\documentclass[twocolumn,floatfix,superscriptaddress,amsmath,amssymb,aps,pra]{revtex4-2}

\pdfoutput=1
\usepackage{graphicx}
\graphicspath{ {Figures/} }

\usepackage{dcolumn}
\usepackage{bm}
\usepackage{bbold}
\usepackage[export]{adjustbox}

\usepackage{multirow}

\usepackage{array,colortbl,xcolor}

\usepackage[utf8]{inputenc}
\usepackage[T1]{fontenc}
\usepackage{mathrsfs,bbm}
\usepackage{dsfont}

\usepackage{amssymb}
\usepackage{amsmath}
\usepackage{amsfonts}
\usepackage{amsthm}
\usepackage{mathtools}
\usepackage{hyperref}
\hypersetup{pdfpagemode=UseNone}
\hypersetup{colorlinks=true}
\hypersetup{citecolor=blue}
\hypersetup{linkcolor=blue}
\hypersetup{urlcolor=blue}
\usepackage{cleveref}

\usepackage{latexsym}
\usepackage{color}
\usepackage{enumerate}
\usepackage{comment}

\newtheorem{proposition}{Proposition}
\newtheorem{theorem}[proposition]{Theorem}
\newtheorem{lemma}[proposition]{Lemma}

\theoremstyle{definition}
\newtheorem{example}[proposition]{Example}
\newtheorem{definition}[proposition]{Definition}

\newcommand{\R}{\mathbb{R}} 
\renewcommand{\nat}{\mathbb N} 

\newcommand{\tr}[1]{{\rm tr}\left[#1\right]} 
\newcommand{\id}{\mathbbm{1}} 

\newcommand{\uleq}{\preceq} 

\newcommand{\Qd}{\mathcal{Q}_d} 
\newcommand{\Cd}{\mathcal{C}_d} 
\newcommand{\dimq}{\mathrm{dim}_\mathcal{Q}}
\newcommand{\dimc}{\mathrm{dim}_\mathcal{C}}

\newcommand{\rank}[1]{\mathrm{rank}(#1)} 
\newcommand{\nrank}[1]{\mathrm{rank}_+(#1)} 
\newcommand{\rnrank}[1]{\mathrm{rank}^*_+(#1)} 
\newcommand{\psdrank}[1]{\mathrm{rank}_{psd}({#1})} 

\newcommand{\ceil}[1]{\lceil {#1} \rceil}

\date{\today}

\begin{document}

\title{Simple Information Processing Tasks with Unbounded Quantum Advantage}

\author{Teiko Heinosaari}
\email{teiko.heinosaari@jyu.fi}
\affiliation{Faculty of Information Technology, University of Jyväskylä, Finland}

\author{Oskari Kerppo}
\email{oskari.e.o.kerppo@jyu.fi}
\affiliation{Faculty of Information Technology, University of Jyväskylä, Finland}

\author{Leevi Lepp\"aj\"arvi}
\email{leevi.leppajarvi@savba.sk}
\affiliation{RCQI, Institute of Physics, Slovak Academy of Sciences,
D\'ubravsk\'a cesta 9, 84511 Bratislava, Slovakia}

\author{Martin Plávala}
\email{martin.plavala@uni-siegen.de}
\affiliation{Naturwissenschaftlich-Technische Fakultät, Universität Siegen, Walter-Flex-Straße 3, 57068 Siegen, Germany}

\begin{abstract}
Communication scenarios between two parties can be implemented by first encoding messages into some states of a physical system which acts as the physical medium of the communication and then decoding the messages by measuring the state of the system. We show that already in the simplest possible scenarios it is possible to detect a definite, unbounded advantage of quantum systems over classical systems. We do this by constructing a family of operationally meaningful communication tasks each of which on one hand can be implemented by using just a single qubit but which on the other hand require unboundedly larger classical system for classical implementation. Furthermore, we show that even though with the additional resource of shared randomness the proposed communication tasks can be implemented by both quantum and classical systems of the same size, the number of coordinated actions needed for the classical implementation also grows unboundedly. In particular, no finite storage can be used to store all the coordinated actions required to implement all possible quantum communication tasks with classical systems. 
As a consequence, shared randomness cannot be viewed as a free resource.
\end{abstract}

\maketitle

\section{Introduction}
One of the most important ongoing investigations in quantum information is to find information processing tasks that can be used to demonstrate a quantum advantage. This question is equally important for quantum computing, where it is still unclear whether the current NISQ devices \cite{preskill2018quantum} can demonstrate a computational advantage over classical computers. Therefore, it is of utmost importance to identify scenarios where quantum devices can offer some benefit over devices operating in the classical domain. Understanding the basic ingredients of quantum advantage forms the foundation for quantum technological applications.

A simple yet important information processing task is the one-way communication scenario, where one party, Alice, is trying to send an encoded message to another party, Bob. The message is prepared with a preparation device, while the encoded message, carried by some physical medium, is decoded with a measurement device. While this scenario may seem far too simple at first to reveal anything interesting, large separations between classical and quantum protocols in such scenarios have been shown in entanglement-based setups \cite{Massar2001Classical, Toner2003Communication}, in the presence of shared randomness \cite{Renner2023Classical,Ambainis2008Quantum,Guha2021quantumadvantage}, as well as in other settings \cite{Perry2015Communication,Gallego2010Device,Chaves2015Device,Saha2020Advantage,Emariau2022Quantum,Buhrman2001Fingerprinting,Galvao2003Substituting,Ambainis2002Dense,Rout2023Unbounded,Patra2023Classical,Halder2022Quantum}. Exponential separations are also known in the one-way communication complexity scenario \cite{Gavinsky2009Exponential, BarYossef2008Exponential, Anshu2017Exponential} where the aim is to compute the value of a multivariate function; interestingly it is also known that the classical-quantum separation cannot be arbitrarily large in these scenarios \cite{Aaronson2005Limitations,Klauck00}. Moreover one can also see any channel-based quantum computation as a preparation and subsequent measurement of a quantum state and the aforementioned results as proofs that quantum computers have an advantage over classical ones when sampling from some conditional probability distributions. These results are especially important in light of boson sampling \cite{aaronson2013computational,gogolin2013boson,Gard2015Introduction,wang2019boson,zhong2020quantum}, which is a non-universal model of quantum computation based on sampling from specific conditional probability distributions.

\begin{figure}
    \centering
    \includegraphics[width=0.6\linewidth]{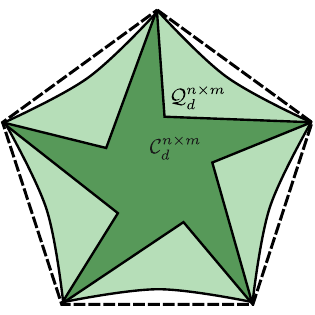}
    \caption{Illustration of the quantum and classical one-way communication scenarios represented by the sets of communication matrices $\Qd^{n \times m}$ and $\Cd^{n \times m}$ of fixed size $n\times m$ and of the same dimension $d$, respectively. The convex closures of both sets are the same (dashed area) although the sets are different.}
    \label{fig:convex-closure}
\end{figure}


A customary assumption in the quantum information literature is that the preparation and measurement devices are classically correlated via shared randomness making the sets of implementable communication scenarios convex and easier to analyze.
Importantly, it is known that in the presence of shared randomness quantum communication does not offer any advantage over classical communication as the sets of implementable scenarios are the same \cite{Frenkel2015Classical}. This hints that either quantum physics does not offer great advantage in one-way communication, or that shared randomness should not be considered as a free resource in these scenarios. In the following we show that the latter is the right conclusion by constructing communication scenarios that require unbounded amounts of shared randomness.

The starting point for our current investigation is the fact that in the absence of shared randomness the sets of implementable communication scenarios are not convex \cite{Gallego2010Device,HeKeLe2020}, see Fig.~\ref{fig:convex-closure} for illustration. It is clear that there can exist an information processing advantage only if the sets of implementable communication scenarios are different between two communication mediums and it has been previously shown that the sets of quantum and classical communication scenarios with systems of fixed dimension are not the same, but they have some rather large discrepancies \cite{HeKeLe2020}. This raises the paramount question if quantum communication can be simulated with a classical system of larger dimension.

In this 
work
we argue that quantum communication cannot be reliably simulated with larger classical systems. This is shown by introducing a family of simple communication scenarios based on antidistinguishability which can be implemented with a qubit, but require arbitrarily large classical systems or arbitrary amount of classical resources in the form of classical communication or shared randomness in the limit. We then argue that the use of shared randomness cannot be considered as a free resource for communication because the difference between the sizes of the quantum and classical systems needed to implement the communication task puts a limit on how much shared randomness is needed to implement quantum communication classically.

\section{Information processing tasks}
In a communication scenario we can envision the following setup: Alice will be made aware of the value of a random variable $a$, while Bob remains unaware of it. Prior to the task, they have the opportunity to meet and establish any strategy they desire. However, once they start, Alice is restricted from freely communicating with Bob. Instead, she is provided with a $d$-dimensional quantum system, or alternatively, a $d$-dimensional classical system. Alice is revealed the value of $a$ and has the liberty to prepare this system in any desired state and hand it over to Bob. Subsequently, Bob's task is to specify the value $a$. The task can be also more general, where Bob is expected to return different values $b_1,\ldots,b_m$ with some probabilities $p_1,\ldots,p_m$ determined by the input $a$. For instance, it could be required that for some inputs Bob erases all information, meaning that he delivers a uniform probability distribution. 
In a computing scenario the setup is similar, but Alice and Bob can be the same person. In this case, there is a function $f$ and Alice's task is to produce $f(a)$ for any input $a$ drawn from some set of possible inputs. The role of a communication channel is taken by a quantum processor.

\begin{figure}
    \centering
    \includegraphics[width=\linewidth]{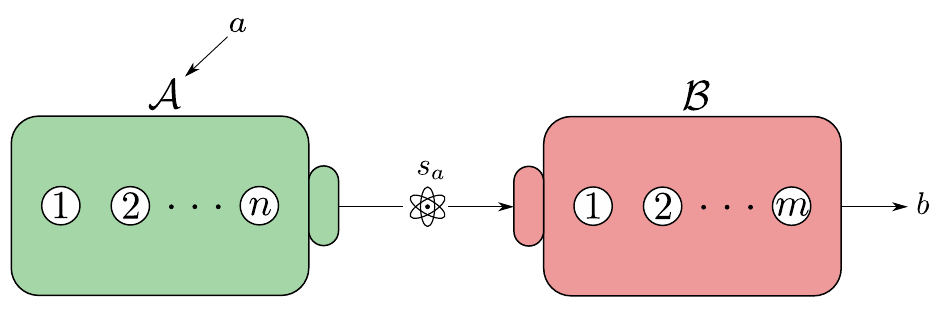}
    \caption{Basic one-way communication setup between two parties, Alice and Bob. On each round of communication Alice receives a random variable $a$ and prepares a state $s_a$ from her state ensemble. Bob performs a measurement of $M$ and receives an outcome $b$.}
    \label{fig:basic-setup}
\end{figure}

To cover all these information processing scenarios, we are using the formalism of communication matrices \cite{Kerppo2,HeKeLe2020}, also called channel matrices \cite{Frenkel2015Classical}. We assume that Alice possesses a finite collection of states, referred to as a state ensemble. Alice selects a label $a$ and transmits a quantum (or classical) system in the corresponding state $s_a$ to Bob. Bob then performs a measurement using a fixed measurement device $M$, which yields $b$, one of the possible outcomes $\{1,\ldots, m\}$. The entire set of conditional probabilities that describes this preparation-measurement scenario is represented by an $n\times m$ matrix $C$, given as $C_{ab}=\tr{s_a M_b}$. This matrix is row-stochastic, i.e., has non-negative entries and the sum of each row is 1. This basic communication scenario is illustrated in Figure \ref{fig:basic-setup}.

The previously mentioned information processing tasks can be written as communication matrices. For instance, the task of perfectly distinguishing $d$ states corresponds to the identity matrix $\id_d$. On the other hand, consider a communication task where there are three possible labels and Alice is able to transmit the first two without error while the third input leads to an ambiguous result. This corresponds to the matrix 
\begin{equation}
\begin{bmatrix}
1 & 0 \\ 0 & 1 \\ \frac{1}{2} & \frac{1}{2} 
\end{bmatrix} \, ,
\end{equation}
where the third row shows that when Alice sends the third label to Bob, Bob gets randomly one of the first labels in his measurement.

To compare the difficulty of implementation of communication matrices corresponding to different tasks, we recall that there is a physically motivated preorder (i.e. partial order without being antisymmetric) called ultraweak matrix majorization in the set of communication matrices \cite{HeKeLe2020}. We write $C\uleq D$ and say a communication matix $C$ is ultraweakly majorized by another communication matrix $D$ if there exist row-stochastic matrices $L,R$ such that $C=LDR$. The matrices $L$ and $R$ have a natural interpretation as pre- and post-processing matrices of the preparations and measurement outcomes: in the communication task $C$ one uses the mixtures of states used in task $D$ according to the convex weights provided by the matrix $L$ and the measurement used in task $C$ is a classical postprocessing given by the matrix $R$ of the measurement used in task $D$. Thus, the preorder gives a precise mathematical definition of the level of difficulty among the tasks: if $C \uleq D$, then performing the task $C$ cannot be any harder than performing $D$ since the implementation of $D$ can be used to implement $C$ as well. In the case when $C \uleq D$ and $D \uleq C$ we say that $C$ and $D$ are ultraweakly equivalent and then they represent equally hard communication tasks.


In the computational scenario a $0/1$-matrix can be interpreted as a Boolean function and we will refer to $0/1$-matrices as deterministic matrices. For instance, the logic gates NOT and XOR correspond to the matrices
\begin{equation} \begin{split}\label{eq:XOR}
\begin{bmatrix}
0 & 1 \\ 1 & 0 
\end{bmatrix} \quad \textrm{and} \quad
\begin{bmatrix}
1 & 0 \\ 0 & 1 \\ 0 & 1 \\ 1 & 0 
\end{bmatrix} \, .
\end{split} \end{equation} 
The NOT setup simply means that Bob reads the opposite label that Alice sends. In the XOR setup Alice can send all ordered pairs from $\{0,1\}$, hence the matrix has four rows. It is straightforward to verify that the communication matrices in \eqref{eq:XOR} are ultraweakly equivalent.

There is a relatively widely adopted view that a single qubit on its own is not more powerful than a bit. This view is motivated by Holevo's theorem \cite{Holevo1973}, which states that a single qubit can only transmit one bit of information. 
Indeed, we make the following observation. 

\begin{proposition}\label{prop:0-1-proposition}
    Let $C$ be a deterministic communication matrix. Then $C$ is implementable with classical and quantum systems of the same size.
\end{proposition}
\begin{proof}
    Let $C$ be a $0/1$-matrix, i.e., a deterministic communication matrix. It is straightforward to see that any communication matrix is ultraweakly equivalent with the matrix that is obtained by removing the zero columns (corresponding to measurement outcomes which never occur) and duplicate rows (some of the states that are used in the implementation are the same). Denote by $C'$ the ultraweakly equivalent matrix that is obtained from $C$ in this way. But now $C'$ is clearly a permutation of an identity matrix of some size $d$, depending on how many duplicate rows were removed. As $C'$ is equivalent to $\id_d$ and the ultraweak relation is clearly transitive, we conclude that $C$ is implementable by $d$-level classical and quantum systems as both of these systems can transmit $d$ distinct messages without error by the basic decoding theorem \cite{Schumacher2010Quantum}.
\end{proof}

By Proposition \ref{prop:0-1-proposition} we are forced to conclude that by looking at deterministic communication matrices, quantum systems do not seem to offer any advantage over classical systems. However, as we will see, this if far from the whole truth as 
there are many interesting and important communication matrices that are not deterministic. In the current framework it becomes natural to look at all possible communication matrices that can be implemented with a $d$-dimensional quantum or classical system.
We denote these sets as $\Qd$ and $\Cd$, respectively.
We note that both of these sets consist of arbitrarily large matrices. Both of them also contain the identity matrix $\id_d$, but not $\id_{n}$ for any $n>d$. 

The sets $\Qd$ and $\Cd$ do not have convex structure as they have matrices of different sizes. To form convex mixtures, we have to limit to matrices of certain size and we denote by $\Qd^{n\times m}$ and $\Cd^{n\times m}$ the matrices with size $n\times m$. 
Forming a convex combination of two communication matrices of the same size corresponds to a coordinated action with Alice and Bob.
As mentioned previously, the sets $\Qd^{n\times m}$ and $\Cd^{n\times m}$ are not convex \cite{Gallego2010Device, Kerppo2023Quantum} but in the seminal work \cite{Frenkel2015Classical} it was shown that the convex closures of $\Qd^{n\times m}$ and $\Cd^{n\times m}$ are the same, the convex structure of these sets is illustrated in Figure \ref{fig:convex-closure}. It follows that shared randomness is a genuinely distinguished resource in the setup but with unlimited use of shared randomness one can simulate any quantum prepare-measure scenario with classical settings. Our main result is to show that this would require infinitely many rounds of coordinated actions.

\section{Classical and quantum dimensions}
To explain the information processing tasks that cannot be effectively simulated with a classical system with the same dimension, we need some additional tools.
The \emph{quantum dimension} of a communication matrix $C$, denoted by $\dimq(C)$, is the smallest integer $d$ such that $C_{ab}=\tr{s_a M_b}$, where the states $\{ s_a \}$ and effects $\{M_b \}$ act on a $d$-dimensional Hilbert space. As the matrix $C$ has finite size, the quantum dimension of a fixed communication matrix is also always finite. Thus, for a communication matrix $C$ we have that $C \in \Qd$ if and only if $\dimq(C) \leq d$. It can be shown that the quantum dimension of a communication matrix $C$ is equal to the positive semidefinite (PSD) rank of $C$ \cite{FiMaPoTiWo12,Lee2017Some}, denoted by $\psdrank{C}$, since a PSD decomposition can always be normalized to give density operators and effects in the decomposition.

The \emph{classical dimension} of a communication matrix $C$, denoted by $\dimc(C)$, is the smallest integer $d$ such that $C$ can be obtained using classical system with $d$ distinct states. Equivalently, it is the smallest integer $d$ such that the implementation of the matrix is obtained by using the standard embedding of classical $d$-dimensional theory in the quantum formalism by using only diagonal density matrices and effects of dimension $d$. It then follows that $C \in \Cd$ if and only if $\dimc(C) \leq d$. Furthermore, it can be shown that $\dimc(C)$ is then the smallest integer $d$ such that there exist row-stochastic matrices $L,R$ such that $C=L\id_d R$, i.e., $C \uleq \id_d$ \cite{HeKeLe2020}. Thus, the communication tasks with a classical implementation of dimension $d$ are exactly those that can be obtained from the task of perfectly distinguishing $d$ states by the means of ultraweak matrix majorization, i.e., in terms of pre- and postprocessing of states and measurement outcomes. From this equivalent formulation it is then clear that mathematically the classical dimension of a communication matrix $C$ exactly coincides with the concept of nonnegative rank of $C$ \cite{Cohen1993Nonnegative}, denoted by $\nrank{C}$.

The connection between the quantum and classical communication complexity and the PSD and nonnegative ranks is known in the literature (see e.g. \cite{FiMaPoTiWo12}). It is also known that these ranks are very difficult to compute \cite{Shitov2016The, Vavasis2009On}. Nontrivial upper bounds on the nonnegative rank have been derived only relatively recently \cite{Gillis2012On, Shitov2014An}.  Our strategy is to define a family of communication matrices with fixed quantum dimension, and show that the classical dimension of these matrices cannot be bounded from above. We note that previously for some matrices (such as the Euclidean distance matrices) an exponential separation between the non-negative and the PSD ranks has been shown \cite{Hrubes12,FaGoPaRoTh15}. However, our aim is to construct actual communication matrices which have a physically meaningful interpretation as clear communication tasks such that they still experience the same exponential separation. To this end, we make the following definition.

\begin{definition}
Let $n \geq 2$ be an integer. Define Bloch vectors $\{ \vec{r}_a \}_{a=1}^n \subset \R^3$ by $\vec{r}_a = \left(
    \cos \left( \frac{2a\pi}{n} \right), 0, \sin \left( \frac{2a\pi}{n} \right)
\right)$, so that $s_a = \frac 12 \left( \id + \vec{r}_a \cdot \vec{\sigma} \right)$, where $\vec{\sigma} = (\sigma_x, \sigma_y, \sigma_z)$ is composed of the Pauli matrices, defines a pure state for all $a$. Define the corresponding effects by $M_b = \frac 1n \left( \id - \vec{r}_b \cdot \vec{\sigma} \right)$, so that $\tr{s_a M_b} = \frac{2}{n}\sin^2 \left(  \frac{(a-b)\pi}{n} \right)$. Note that $\tr{s_c M_c} = 0$ for all $c$ and otherwise $\tr{s_a M_b} > 0$. We denote with $A_n$ the $n \times n$ matrix with elements $\tr{s_a M_b}$ at the $a$th row and $b$th column. 
\end{definition}


The matrix $A_n$ defines an instance of communication tasks known as antidistinguishability \cite{Heinosaari2018Antidistinguishability, Caves2002Conditions, Bandyopadhyay2014Conclusive, Havlicek2020Simple, russo2023inner}. As all diagonal entries of $A_n$ are zero, it corresponds to the task where upon obtaining outcome $b$ Bob knows that Alice did not send state $s_b$. Antidistinguishability plays an important role in quantum information and foundations as evidenced by its role in the influential PBR \cite{Pusey2012On} theorem and its connection to noncontextuality inequalities \cite{Leifer2020Noncontextuality}. Our motivation in defining $A_n$ is that clearly this matrix has quantum dimension equal to two for any $n$. We now proceed to show that the classical dimension of $A_n$ cannot be bounded from above.

\begin{lemma}\label{lemma:nonnegative-scaling}
The nonnegative rank of $A_n$ scales as $\nrank{A_n} = \Omega (\log n)$.
\end{lemma}
The result follows from Theorems 5, 6 and 8 of \cite{Gillis2012On}.  The details can be found in Appendix \ref{appA}. We now have all necessary tools to prove one of our main results.

\begin{theorem}\label{theorem:1}
For all integers $d,m \geq 2$ there is a communication matrix $C$ such that $\dimq(C)\leq d$ while $\dimc(C)\geq m$. 
\end{theorem}
\begin{proof}
    It is sufficient to consider $d=2$. Let $m\geq 2$ and $n=2^m$. It is guaranteed by Lemma \ref{lemma:nonnegative-scaling} that $\dimc(A_n)=\nrank{A_n}\geq m$.
\end{proof}

Our result shows that there is a sequence of communication tasks that can be implemented with a fixed-size quantum system but which cannot be implemented by any  classical system of fixed-size. In the limit of these tasks we arrive at an unbounded quantum advantage over any classical implementation. In the literature quantum advantages are typically shown in scenarios involving inputs for both parties of the communication protocol. In the present work we study the scenario where only the party preparing states receives an input and we were thus able to show the advantage with communication tasks of the simplest type with a very clear and important physical motivation, namely antidistinguishability of states. This result adds to the list of topics how antidistinguishability can be used to surface some of the fundamental non-classical aspects of quantum theory as well as what type of information processing capabilities it holds. On the other hand, our result can also be used as a witness for the classical dimension of the implementation: if we know that we are using a classical system to implement an antidistiguishability task of certain size, then we can use the lower bound from Lemma \ref{lemma:nonnegative-scaling} to deduce what the dimension of the system must at least be. Additionally, we consider it important that we managed to use preparations and measurements that belong only to $XZ$-plane of the Bloch sphere, since this immediately shows that our construction is also valid for real quantum theory (and also for PSD rank based on real Hilbert spaces).

From the technical side a natural question is whether it is possible to improve the scaling of the nonnegative rank in Lemma \ref{lemma:nonnegative-scaling}? A nontrivial upper bound on the nonnegative rank of $\ceil{\frac{6n}{7}}$ was proven in \cite{Shitov2014An}. 
Therefore better scaling could be possible but we leave this as an open problem. 

\begin{example}\label{ex:A_7}
$A_7$ is the matrix with elements $(A_7)_{ab} = \frac{2}{7}\sin^2 \left(  \frac{(a-b)\pi}{7} \right)$ with $a,b \in \{1,2,\dots,7 \}$. Define $i=\frac 27 \sin^2\left( \frac{\pi}{7} \right)$, $j=\frac 27 \sin^2\left( \frac{2\pi}{7} \right)$ and $k=\frac 27 \sin^2\left( \frac{3\pi}{7} \right)$. Then 
\begin{equation} 
\begin{split}
    A_7 = \begin{bmatrix}
        0 & i & j & k & k & j & i \\
        i & 0 & i & j & k & k & j \\
        j & i & 0 & i & j & k & k \\
        k & j & i & 0 & i & j & k \\
        k & k & j & i & 0 & i & j \\
        j & k & k & j & i & 0 & i \\
        i & j & k & k & j & i & 0 
    \end{bmatrix}.
\end{split} 
\end{equation}
$A_7$ has the following nonnegative factorization as $A_7 = WH$, where
\begin{equation} 
\begin{split}
    W &= \begin{bmatrix}
        2k & 2j & 0 & 0 & 2i & 0 \\
        0 & 2k & 0 & 0 & \frac{2ik}{k-j} & w \\
        0 & 2j & 2k & 0 & 2i & 0 \\
        0 & 2i & \frac{2k(j-i)}{k-j} & 1-RS^* & 0 & 0 \\
        0 & 0 &  \frac{2ik}{k-j} & 2(k-\frac{i^2}{k-j}) & 1-RS^* & w \\
        \frac{2ik}{k-j} & 0 & 0 & 2(k-\frac{i^2}{k-j}) & 1-RS^* & w\\
        \frac{2k(j-i)}{k-j} & 2i & 0 & 1-RS^* & 0 & 0
    \end{bmatrix}, \\
    H &=\begin{bmatrix}
        0 & \frac{i}{2k} & \frac{j}{2k} & \frac{k-i}{2k} & \frac{k-j}{2k} & 0 & 0 \\
        0 & 0 & 0 & 0 & \frac 12 & \frac 12 & 0 \\
        \frac{j}{2k} & \frac{i}{2k} & 0 & 0 & 0 & \frac{k-j}{2k} & \frac{k-i}{2k}\\
        \frac 14 & \frac 12 & \frac 14 & 0 & 0 & 0 & 0 \\
        0 & 0 & 0 & \frac 12 & 0 & 0 & \frac 12 \\
        h_1 & 0 & h_1 & h_2 & 0 & 0 & h_2
    \end{bmatrix},
\end{split}
\end{equation} 
where 
\begin{equation} \begin{split}
w&=2(i+j)-\dfrac{2ik}{k-j}, \quad
h_1=\dfrac{j-\frac 12 (k-\frac{i^2}{k-j})}{2(i+j)-\frac{2ik}{k-j}}, \\
h_2&=\dfrac{j-\frac{ik}{k-j}}{2(i+j)-\frac{2ik}{k-j}},
\end{split} \end{equation} 
and $RS^*$ stands for the sum of all other elements on the same row. This nonnegative factorization was found with the help of a heuristic method \cite{Vandaele2016Heuristics}. As this decomposition exists, the nonnegative rank of $A_7$ is less than or equal to six. The bound given in \cite[Corollary 4]{Gillis2012On} can be used to show that equality must hold. The details can be found in Appendix \ref{appB}.

\end{example}

The classical dimension of $A_7$ provides a counterexample to a conjecture presented in \cite{HeKeLe2020} where it was speculated that the classical dimension of quantum theory equals the quantum dimension squared.

\section{Nonnegative rank and shared randomness}
The first part of our result was to show that the classical dimension of communication matrices with fixed quantum dimension can grow without bound, we now turn to our second result concerning shared randomness. Namely, it is well-known that any quantum communication matrix can be obtained using classical communication and shared randomness \cite{Frenkel2015Classical}. Since shared randomness is usually considered a free resource it would seem that our previous result holds no real consequence and shared randomness could be used as a loophole. Next we will close this loophole by showing that shared randomness is not in fact a free resource.

Suppose Alice and Bob coordinate their actions in the following way: Alice and Bob both have $k \in \nat$ different choices for their preparation and measurement devices. That is, in each round of communication, Alice and Bob observe a correlated random variable $k' \in \{1,2,\dots,k\}$. Alice prepares some state with the preparation device labeled $k'$. Likewise Bob uses measurement device with label $k'$. Without loss of generality we can assume that the devices have the same number of inputs and outputs, say $n$ preparations and $m$ outcomes, so that the coordinated communication matrix can be written as 
\begin{equation} \begin{split}
    C = \sum_{k'=1}^k \alpha_{k'} C_{k'},
\end{split} \end{equation}  
where $\alpha_{k'}$ is the probability of sampling $k'$ as the shared variable and $C_{k'}$ is the corresponding communication matrix implemented by those devices. Suppose now that each of the $C_i$'s has classical dimension equal to $d$, that is, $C_{k'} = L_{k'} R_{k'}$ for some row-stochastic matrices $L_{k'}$ and $R_{k'}$ of size $n \times d$ and $d \times m$, respectively. Then clearly \begin{equation} \begin{split}\label{eq:SR-bound}
    C = \begin{bmatrix}
        \alpha_1 L_1 & \alpha_2 L_2 & \dots & \alpha_k L_k
    \end{bmatrix}\begin{bmatrix}
        R_1 \\ R_2 \\ \vdots \\ R_k
    \end{bmatrix}.
\end{split} \end{equation} 
This puts an upper bound on the classical dimension of $C$, namely, $\nrank{C}\leq dk$.

\begin{theorem}
    Suppose Alice and Bob agree beforehand that they will use coordinated actions from a finite set $\{1,2,\dots ,k\}$ and they will communicate with $d$-dimensional classical systems. There exists communication matrices with quantum dimension equal to two that they cannot implement classically by coordinating their actions.
\end{theorem}
\begin{proof}
    The result follows directly from Equation \eqref{eq:SR-bound} and Theorem \ref{theorem:1}, as the maximal classical dimension Alice and Bob can achieve is $dk$ and we have seen in Theorem \ref{theorem:1} that the classical dimension of antidistinguishability matrices cannot be bounded.
\end{proof}

Our second main result thus shows that ultimately shared randomness cannot be considered a free resource. In particular, even though the convex closures of $\Cd$ and $\Qd$ are the same for fixed $d$ \cite{Frenkel2015Classical}, one cannot implement all the tasks in $\Qd$ by using a $d$-dimensional classical system with any finite amount of shared randomness. In the limit this ultimately leads to the need for an infinite storage for the coordinated actions. It is worth repeating that without this result our first result would lack proper physical and operational interpretation and hence it is a key feature that this loophole was closed. Furthermore, there are no further known loopholes remaining in this scenario. On the other hand, our second result can also be used as a witness for shared randomness: there are simple communication tasks which for fixed dimension $d$ can be used to detect how many coordinated actions is minimally needed to implement a given task classically. 

\section{Discussion}
We have shown that it is not possible to simulate quantum communication with larger classical systems reliably. Our result has two consequences. First of all, if the dimension of the communication medium is taken to be proportional to the cost of communication, then quantum communication displays an unbounded advantage. As quantum systems can only give an advantage in the considered simple communication scenarios when the communication matrix is not deterministic, we conclude that the displayed advantage is somehow related to quantum systems generating randomness more effectively than is possible classically. In the example we provided with antidistinguishable matrices the classical dimension scaled at least logarithmically with the size of the matrix. We are unsure if it is possible to obtain a better scaling with other matrices, or even what the exact scaling of the antidistinguishable matrices are. We leave this problem for future work.

Secondly, given that the convex closures of quantum and classical communication matrices coincides, we nonetheless have shown that no finite amount of coordinated actions is enough to exactly simulate quantum communication classically when the preparation and measurement devices are correlated. This points us to the conclusion that taking shared randomness as a free resource is an overwhelmingly strong assumption.

\begin{acknowledgments}

TH and OK acknowledge financial support from the Business Finland under the project TORQS, Grant 8582/31/2022, and from the Academy of Finland under the mobility funding Grant No. 343228, and under the project DEQSE, Grant No. 349945.

LL acknowledges support from the European Union’s Horizon 2020 Research and Innovation Programme under the Programme SASPRO 2 COFUND Marie Sklodowska-Curie grant agreement No. 945478 as well as from projects APVV-22-0570 (DeQHOST) and VEGA 2/0183/21 (DESCOM).

MP acknowledges support from the Deutsche Forschungsgemeinschaft (DFG, German Research Foundation, project numbers 447948357 and 440958198), the Sino-German Center for Research Promotion (Project M-0294), the ERC (Consolidator Grant 683107/TempoQ), the German Ministry of Education and Research (Project QuKuK, BMBF Grant No. 16KIS1618K), and the Alexander von Humboldt Foundation.
\end{acknowledgments}

\bibliography{bibliography}

\newpage

\onecolumngrid


\appendix


\section{Proof of Lemma \ref{lemma:nonnegative-scaling}}\label{appA}

The restricted nonnegative rank of a nonnegative matrix $A$, denoted by $\rnrank{A}$, is defined as the smallest decomposition of $A$ into $A=WH$ with $W,H$ nonnegative and $\rank{A} = \rank{W}$. Clearly, if $A$ is of size $n \times m$, then $\rank{A} \leq \nrank{A} \leq \rnrank{A} \leq m$. The authors in \cite{Gillis2012On} show that the restricted nonnegative rank is related to the nested polytopes problem which they can use to show bounds on the restricted nonnegative rank and the nonnegative rank. In particular, they give the following upper bound on the restricted nonnegative rank.

\begin{theorem}[\cite{Gillis2012On}, Theorem 5]\label{thm:Gillis-ub-rnrank}
The restricted nonnegative rank of a nonnegative matrix $M$ with $r = \rank{M}$ and $r_+ = \nrank{M}$ can be bounded above by
\begin{equation} \begin{split}
\rnrank{M} \leq \max_{r\leq r_u \leq r_+}  \mathrm{faces}(r_+,r_u-1, r_u-r) =: \phi_r(r_+),
\end{split} \end{equation} 
where $\mathrm{faces}(n, d, k)$ is the maximal number of $k$-faces of a polytope with $n$ vertices in dimension $d$ and which can be calculated as 
\begin{equation} \begin{split}
\mathrm{faces}(n,d,k) 
&= \sum_{i=0}^{\frac d2} {}^*\left(\begin{pmatrix}d-i \\ k+1-i\end{pmatrix}+\begin{pmatrix}
        i \\ k+1-d+i
    \end{pmatrix}\right)    
    \begin{pmatrix}
        n-d-1+i\\i
    \end{pmatrix}
\end{split} \end{equation} 
and $\sum {}^*$ stands for a sum where half of the last term is taken for $i=\frac{d}{2}$ if $d$ is even and the whole last term is taken when $d$ is odd, or $i = \frac{d-1}{2}$.
\end{theorem}

Furthermore, they show that in fact the quantity $\phi_r(r_+)$ can be used as a lower bound for the nonnegative rank:

\begin{theorem}[\cite{Gillis2012On}, Theorem 6]\label{thm:Gillis-lb-nrank}
The upper bound $\phi_r(r_+)$ on the restricted nonnegative rank of a nonnegative matrix M with with $r = \rank{M}$ and $r_+ = \nrank{M}$ satisfies
\begin{equation} \begin{split}
\phi_r(r_+)= \max_{r\leq r_u \leq r_+}  \mathrm{faces}(r_+,r_u-1, r_u-r) \leq \max_{r \leq r_u \leq r_+} \begin{pmatrix}r_+ \\ r_u-r+1 \end{pmatrix} \leq \begin{pmatrix}r_+ \\ \lfloor r_+/2 \rfloor \end{pmatrix} \leq 2^{r_+} \sqrt{\dfrac{2}{\pi r_+}} \leq 2^{r_+}\, .
\end{split} \end{equation} 
\end{theorem}

By combining Theorems \ref{thm:Gillis-ub-rnrank} and \ref{thm:Gillis-lb-nrank} we see that for a nonnegative matrix $M$ it holds that 
\begin{equation}\label{eq:thm5&6}
\rnrank{M} \leq 2^{\nrank{M}}\, .
\end{equation}
Thus, the restricted nonnegative rank gives a lower bound on the actual nonnegative rank. In order to use this bound, we must be first able to calculate (or lower-bound) the restricted nonnegative rank. For a particular class of matrices, which will also suit our purposes, the authors in \cite{Gillis2012On} were able to calculate the restricted nonnegative rank exactly.

Let $A^i$ denote the $i$th row of a matrix $A$. The sparsity pattern of a row $A^i$ is defined as $S_i = \{ k \, | \, A_{ik} = 0 \}$. The matrix $A$ is said to have a disjoint sparsity pattern if $S_i \nsubseteq S_j$ for all $i \neq j$.  

\begin{theorem}[\cite{Gillis2012On}, Theorem 8]\label{thm:Gillis-rnrank-rank3}
If $M$ is a rank-three nonnegative square matrix of dimension $n$ whose columns have disjoint sparsity patterns, then $\rnrank{M} =n$.
\end{theorem}

Now we can use all the previously stated results of \cite{Gillis2012On} to prove a lower bound for the antidistinguishability matrices $A_n$ from Example \ref{ex:A_7}.
\newtheorem*{lemma1}{Lemma \ref{lemma:nonnegative-scaling}}
\begin{lemma1}
The nonnegative rank of $A_n$ scales as $\nrank{A_n} = \Omega (\log n)$.
\end{lemma1}
\begin{proof}
We note that trivially $\nrank{A_1} = 1$ and $\nrank{A_2}=2$ so that for $n=1$ and $n=2$ the result holds. On the other hand, since the communication matrix $A_n$ can be implemented with a restricted qubit (the states are only in the xz-plane on the Bloch sphere), it can be shown that the maximum rank of these matrices can be at most three \cite[Proposition 3]{HeKeLe2020}. Furthermore it can be easily checked that for matrix size of more than two the rank cannot be two. Thus, it follows that $\rank{A_n}=3$ for all $n > 2$.  Thus, for $n>2$ we have that $A_n$ is a rank-three matrix which clearly has disjoint sparsity pattern so that by Theorem \ref{thm:Gillis-rnrank-rank3} we have that $\rnrank{A_n}=n$. By using the lower bound for the nonnegative rank given by Theorems \ref{thm:Gillis-lb-nrank} and \ref{thm:Gillis-ub-rnrank}, namely Eq. \eqref{eq:thm5&6}, we have that $\nrank{A_n} \geq \log_2 n$ which completes the proof.
\end{proof}

\section{Proof that $\nrank{A_7} =6$} \label{appB}

Since by Theorem \ref{thm:Gillis-rnrank-rank3} we know that $\rnrank{A_7} = 7$, we can use the derived upperbounds from  \cite{Gillis2012On} for the restricted nonnegative rank based on the actual nonnegative rank to lower bound the nonnegative rank of $A_7$. In particular, one could use Theorem \ref{thm:Gillis-ub-rnrank} for this but in fact for rank-three matrices the authors in \cite{Gillis2012On} provide an even improved bound:

\begin{lemma}[\cite{Gillis2012On}, Corollary 4]\label{lemma:restricted-rank-upper-bound}
For a rank-three nonnegative matrix $M$ with $r_+ = \nrank{M}$ the restricted nonnegative rank is bounded above by 
\begin{equation} \begin{split}
\rnrank{M} \leq \max_{3\leq r_u \leq r_+} \min_{i=0,1} \mathrm{faces}(r_+,r_u-1, r_u-3+i) =: \phi'(r_+) \leq \phi_3(r_+) \, .
\end{split} \end{equation} 
\end{lemma}

\begin{table}
\caption{First (relevant) values of the upper bounds for the restricted nonnegative rank given in \cite{Gillis2012On}.}
\begin{ruledtabular}
\begin{tabular}{c c c c c c}
$r_+$ &  3 & 4 & 5 & 6 & 7 \\
$\phi'(r_+)$ & 3 & 4 & 6 & 9 & 14 \\
$\phi_3(r_+)$ & 3 & 6 & 10 & 18 & 30 \\
\end{tabular}
\end{ruledtabular}
\label{table:1}
\end{table}

It is crucial to note that $\phi'(r_+)$ (also $\phi_r(r_+)$ for any fixed $r$) is an increasing funtion of its argument $r_+$. This is because $\mathrm{faces}(n,d,k)$ increases with increasing $n$. The first few (relevant) values of $\phi'$ (and $\phi_3$ for comparison) are presented in the Table~\ref{table:1}. In particular, for $A_7$ we have that $\rnrank{A_7}=7$, so that $\phi'(r_+) \geq 7$. From Table  \ref{table:1} we see that then we must have that $\nrank{A_7} \geq 6$. Our explicit nonnegative factorization from Example \ref{ex:A_7} then shows that in fact $\nrank{A_7} = 6$.

\end{document}